# Scaling of Charged particle multiplicity in $^{28}$Si-nucleus interactions

N. Ahmad*

Department of Physics, Aligarh Muslim University, Aligarh-202002 Uttar Pradesh

**Summary**: Multiplicity distributions and their scaling behaviour for various types of secondary charged particles produced in $^{28}$Si-nucleus collisions at 4.5A and 14.5A GeV are investigated. The validity of KNO scaling and its generalized form has been tested by studying the parameters, $S_n(z)$ and $S_n(z')$. The data obey KNO-G scaling for lower multiplicity events and departs from the predictions at the tails of the distributions.

E-mail: nazeer_ahmad_na@rediffmail.com

Introduction:

The main aim of studying relativistic heavy-ion collisions is to understand the properties of the densest and hottest form of the matter. In particular, one can expect to reach the conditions under which phase transition from hadronic matter to quark-gluon plasma (QGP) might take place. Such phase transition could produce fluctuations in phase space. Thus, at present, most of the studies carried out have focused attention primarily on the search of QGP. To be able to search for the QGP production, a deep and clear understanding of the background is required so that relevant information regarding QGP signals may be obtained. However, a number of interesting aspects of the particles produced in these collisions may give vital information about mechanism of particle production in relativistic nuclear collisions.

Several[1-3] attempts were made to predict the multiplicity distribution of particles produced in hadron-hadron (h-h), hadron-nucleus (h-A) and nucleus–nucleus (A-A) collisions. Koba-Nielsen-Olesen (KNO) scaling has been dominant framework to study the behaviour of multiplicity distribution of charged particles produced in high-energy collisions. In view of the Wroblewski relation[2], which has been observed to be in fine agreement with the experimental results, the KNO scaling function, $\psi(z)$ and z have been modified. An attempt is made here to examine the scaling behaviour of the multiplicity distributions of charged secondary particles by studying the scaling parameters like $\psi(z)$, $S_n(z)$ and their modified versions $\psi(z')$, $S_n(z')$ and $R(z)$. For the purpose, data on 4.5A and 14.5A GeV energies in $^{28}$Si-nucleus collisions are analyzed.

Mathematical formalism:

The Koba-Nielsen-Olesen (KNO) scaling function used to describe and predict the multiplicity distribution at a given energy may hold following relations:

$$P_n = \frac{1}{<n>}\psi(z), <n> = \sum_n nP_n \text{ and } z = \frac{n}{<n>} \qquad (1)$$

where n is the number of secondary produced in an interaction, $P_n$ is the probability of producing n secondary in the final state and $\psi(z)$ is energy independent function which obeys following relation:

$$\int_0^\infty \psi(z)dz = 1 \qquad (2)$$

The validity of the KNO scaling function has been tested by studying, $S_n(z) = \sum_{i=n}^{\infty} P_i$ by the earlier workers[5–8]. They showed that the data obey approximately the KNO predictions. The approximation used to explain the deviations were not found perfect. Therefore, a great deal of effort was done to generalize KNO scaling by Golokhvastov[9] to find a way of describing the multiplicity data in a full energy range by single scaling function. The generalized KNO function, KNO-G is continuous function of a continuous parameter, $n'$ and is related with the probability, $P_{n'}$ is the following fashion:

$$P_n = \int_n^{n+1} P(n')dn' = \int_{n/<n'>}^{n+1/<n'>} \psi(z')dz' \text{ and } P_n = \frac{1}{<n'>}\psi\left(\frac{n'}{<n'>}\right) \qquad (3)$$

where $<n'> \approx <n> + 0.5$ and $z' = \frac{n'}{<n'>}$.

In order to test the validity of the KNO-G scaling function the variation of the parameter, $S_n(z')$ where $S_n(z') = \sum_{i=n}^{\infty} P_i$, with $z'\left(=\frac{n'}{<n'>}\right)$ is studied for the different projectile energies. KNO-G predictions may also be tested by examining whether an individual multiplicity distribution can be reproduced by the scaling function, $\phi(z')$ which obey following relations:

$$\phi(z') = \int_0^{z'} \psi(z')dz' \text{ and } P_n^{th}(n') = \phi_\circ\left(\frac{n+1}{<n'>}\right) - \phi_\circ\left(\frac{n}{<n'>}\right) \qquad (4)$$

The scaling behaviour of relativistic charged particle multiplicity is described by Adamovic et al.[10–12] and was found that the ratio of the density of charged particles with fixed n, $\rho_n$ to the mean density, $<\rho>$ depends only on the scaling variable, z and is insensitive on the projectile beam energy such that:

$$R(z) = \frac{\rho_n}{<\rho>} = \text{Constant (s)}, \tag{5}$$

where $\sqrt{s}$ is the C.O.M. energy of the projectile beam.

Results and discussion:

Multiplicity distributions of secondary charged particles is one of the most sensitive characteristics of relativistic nuclear collisions for testing the predictions of various theoretical models put forward to explain the mechanism of multiparticle production in heavy-ion collisions. Shown in Fig.1 is the multiplicity distribution of various secondary particles produced in $^{28}$Si-nucleus collisions at 4.5A and 14.5A GeV energies. It is interesting to note from these figures that $n_b$ and $n_g$-distributions having are essentially independent of the projectile beam energy. On the other hand, $n_s$-, $n_c$- and $n_h$- distributions appreciably changes with increasing projectile energy.

Multiplicity distribution of various types of charged secondary particles in a KNO scaling form for 4.5A and 14.5 GeV energies in $^{28}$Si-nucleus interactions are studied and results are displayed in Fig.2. It may be of interest to mention that the $n_g$-, $n_s$-, $n_c$- and $n_h$ distributions may be described by the KNO scaling function, $\psi(z) = <n> P_n$. The result of the experimental data for $n_b$-distribution does not show scaling behaviour. However, a significant departure of the results obtained for the experimental data from KNO predictions are found at higher multiplicity values in the considered interactions. This fact is clear from the figure 2 as all data points at lower multiplicity values almost lie on the curves (Eq. 6) and the scattering only seen at the tails of the multiplicity distributions where the largest experimental (statistical) errors are expected. The solid and dashed curves on the plots are the best fits to the experimental data of the following form:

$$\psi(z) = A * z * e^{-(B*z)} \tag{6}$$

where A and B are the constants. The values of these parameters along with their corresponding $\chi^2$/D.F. obtained for the best fits to the data for the considered interactions are presented in the Table 1. It is worthnoting from the table that parameter A and B are found to be almost independent of the projectile energy for all types of charged particles multiplicity distributions considered in the present study and are nearly equal to corresponding their values $11.005 \pm 0.038$ and $2.232 \pm 0.076$ reported by earlier workers[13–15]. Authors have also found that if the experimental data points from the tails of the multiplicity distributions are not considered due to their less significance the $\chi^2$/D.F reduces to $\cong 0.08$.

Table 1: Values of the parameter A and B appearing in Eq. (6) various types of secondary charged particles produced in $^{28}$Si-nucleus interactions.

| Data type ↓ | | Parameters ↓ | | $\chi^2/D.F.$ |
|---|---|---|---|---|
| | | A | B | |
| $n_b$ | 4.5A GeV | 9.771 ± 1.043 | 2.661 ± 0.161 | 0.11 |
| | 14.5A GeV | 9.904 ± 1.059 | 2.796 ± 0.173 | 0.43 |
| $n_g$ | 4.5A GeV | 12.089 ± 1.704 | 3.059 ± 0.245 | 0.18 |
| | 14.5A GeV | 13.471 ± 1.060 | 3.521 ± 0.237 | 0.23 |
| $n_s$ | 4.5A GeV | 14.763 ± 2.310 | 3.134 ± 0.251 | 0.18 |
| | 14.5A GeV | 15.658 ± 1.525 | 3.661 ± 0.358 | 0.67 |
| $n_c$ | 4.5A GeV | 9.518 ± 0.698 | 2.344 ± 0.098 | 0.19 |
| | 14.5A GeV | 9.387 ± 0.844 | 2.842 ± 0.155 | 0.17 |
| $n_h$ | 4.5A GeV | 11.104 ± 1.229 | 2.971 ± 0.175 | 0.22 |
| | 14.5A GeV | 12.080 ± 1.305 | 3.119 ± 0.213 | 0.24 |

Fig. 3 show in a linear scale, the test of the KNO scaling for different types of the secondary particles at two projectile beams energies. A weak projectile energy dependent in the variations of parameter, $S_n(z)$ with z for various types of secondary particles produced in $^{28}$Si-nucleus interactions. It may be of interest to mention that, significant departure of the experimental results obtained for the hadron-nucleus and nucleus-nucleus collisions from the predicted one has been observed [13–14]. However, in the terms of redefined KNO function (KNO-G), $\psi(z')$ and variable $z'$, data in the limited range of incident energy are reported [7,9] to be reproduced quite well. The results of the improved KNO (KNO-G) scaling function, $\psi(z') = <n'> P_n$ for the same data sets are shown in Fig.4. It may be noted from these figures that all data points almost lie on a single curve (Eq.4) for all types of charged particle probability distributions at the both of the projectile energies. The curves in the figures are the best fits to the experimental data of the Eq. 4.

In order to test the validity KNO-G scaling function, $\psi(z')$, parameter $S_n(z')$ also plotted against $z'$ for $n_b$, $n_g$, $n_s$, $n_c$ and $n_h$-distributions at 4.5A and 14.5 GeV in $^{28}$Si-nucleus interactions are exhibited in Fig.5. The data for the parameter, $S_n(z')$ at considered projectile energies are found to overlap to each other. Experimental results differ from KNO-G predictions at larger multiplicity tails. It is worthmentioning that usual particle production mechanism is responsible for the KNO-G scaling and the deviations of the results from the KNO predictions give an evidence for an extra processes contributing to the high multiplicities.

The scaling behaviour of relativistic charged particle multiplicity have also been studied by investigating the dependence of the ratio of the density of particles, $\rho_n$ in given pseudorapidity windows with fixed number of charged secondaries, n over mean density, $<\rho>$, on the scaling variable, $z$. A stronger scaling of multiplicity was proposed [11,12] at R (z) = z for nucleus-nucleus interactions at high energies. Fig.6 shows the dependence of $R(z)$ on z for two projectile energies. The dependence of the parameter R (z) on z is found to be linear. It is interesting to mention that the relation $R(z) = \rho_n / <\rho>$ is valid, within the limits of experimental errors and is found to be independent of the incident beam energy for the whole interval of z. The straight line in the figures is the least squares fit to the data of the equation having following form:

$$R(z) = \frac{\rho_n}{<\rho>} = a * z + b \tag{7}$$

where a and b are determined by the least squares fit. The values of parameters "a" and "b" are predicted[10] to be 1 and 0 respectively. It is interesting to mention that the observed values of the parameters "a" and "b" are $1.248 \pm 0.100$ and $-0146 \pm 0.054$ and $1.143 \pm 0.095$ and $0.060 \pm 0.092$ for 4.5A and 14.5A GeV in $^{28}$Si-nucleus interactions respectively. It may also be pointed that the observed values of the parameters are found to be essentially independent to the projectile energy. However, a deviation of the R (z) versus z from linearity at higher value of z may be attributed to the fact of some kind of violation of the scaling.

## Conclusions:

On the basis of the present study, we conclude the following:

1. $n_b$ and $n_g$ -distributions are found to be almost independent of the projectile beam energy. However, a weak dependence of, $n_s$- , $n_h$- and $n_c$-distributions on projectile energy is observed.

2. The KNO scaling derived for asymptotic energies is applied to our experimental data. The validity of KNO scaling is tested by studying the variations of $S_n(z)$ with z for various types of secondary particles produced in $^{28}$Si-nucleus interactions at 4.5A and 14.5A GeV. The data almost overlap for both the projectile energies at lower multiplicities values.

3. The KNO-G scaling is tested by studying the variations $S_n(z')$ with $z'$ for various types of secondary charged particles produced in $^{28}$Si-nucleus interactions and is found to be essentially independent of the projectile beam energy. However, data are more scattered from the KNO-G predictions at larger values of multiplicities.

4. Finally, it may be remarked that the scaling violations are small and our experimental data exhibits scaling behaviour within the limit of experimental errors.

**References:**

1. P. Slattery: Phys. Rev. Letter **29** 1624(1972); Phys. Rev. **D7** 2073(1972).
2. P. Olesen: Phys. Lett. **B47** 251(1973).
3. A. Shakeel et al: Physica Scripta **29** 435(1984).
4. A. Break Stone et al: Phys. Rev. **D30** 528(1984).
5. E. D. Wolf et al: Nucl. Phys. **B87** 325 (1975).
6. W. Thome et al: Nucl. Phys. **B129** 356 (1977).
7. A. I. Golokhvastov: Sov. J. Nucl. Phys. **27** 430 (1978); **30** 128(1979).
8. R. Szweed and G, Worehna: Z. Phys. **C29** 255(1985).
9. A. I. Golokhvastov: Sov. J. Nucl. Phys. **27** 430 (1978)
10. O. Adamovic et al: Phys. Rev. **C48** 2772 (1993).
11. G. J. Alner et al: Z. Phys. **C33** 1(1986).
12. S. Backovic et al : Z. Phys. **C53** 613(1992)
13. Alma Ata and et al (Collaboration): Sov. J. Nucl. Phys. **30** 824(1986).
14. A. Shakeel: Ph. D. Thesis submitted to Aligarh Muslim University, Aligarh (1986).

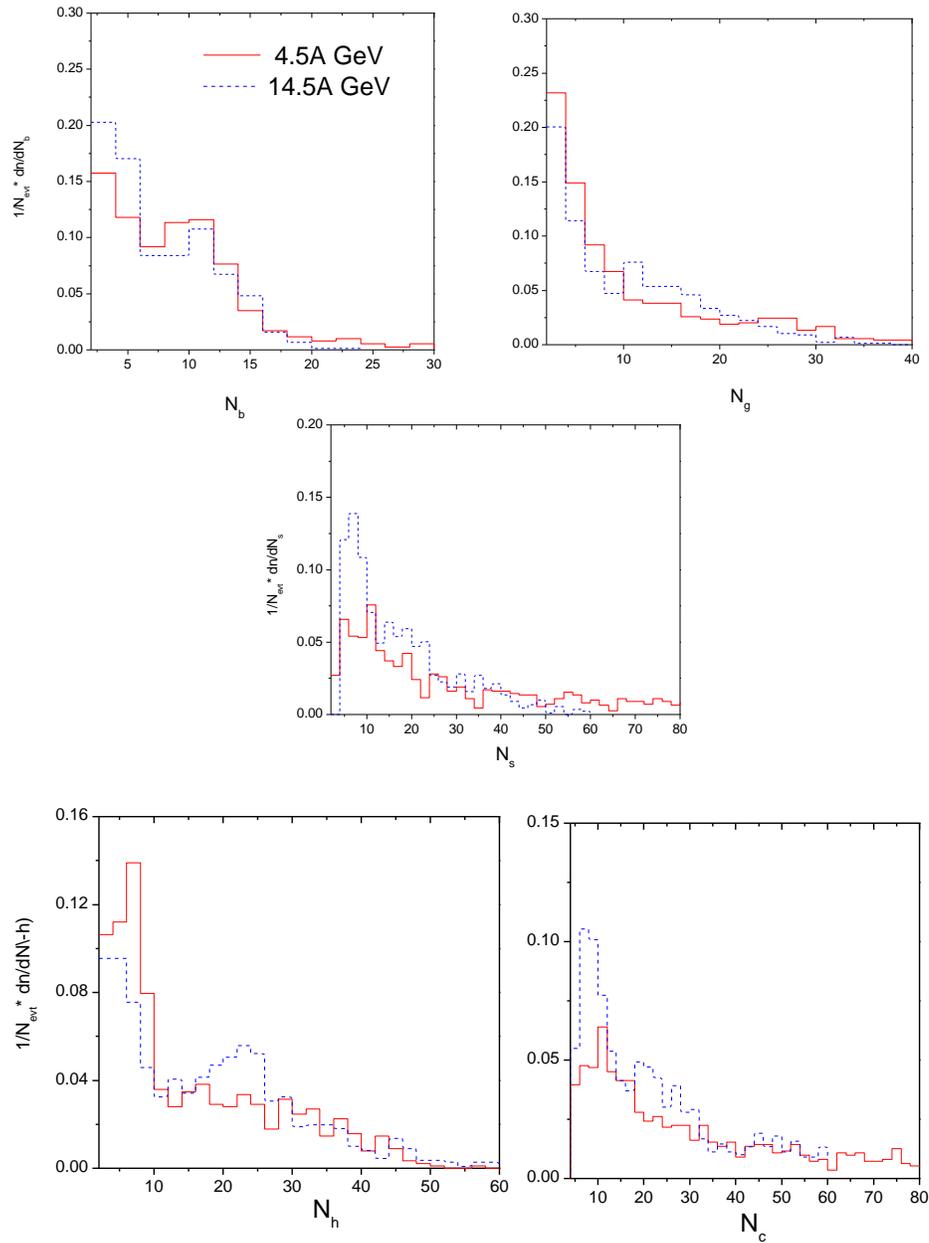

**FIG. 1** Multiplicity distribution of various types of secondary particles produced in $^{28}$Si-nucleus interactions.

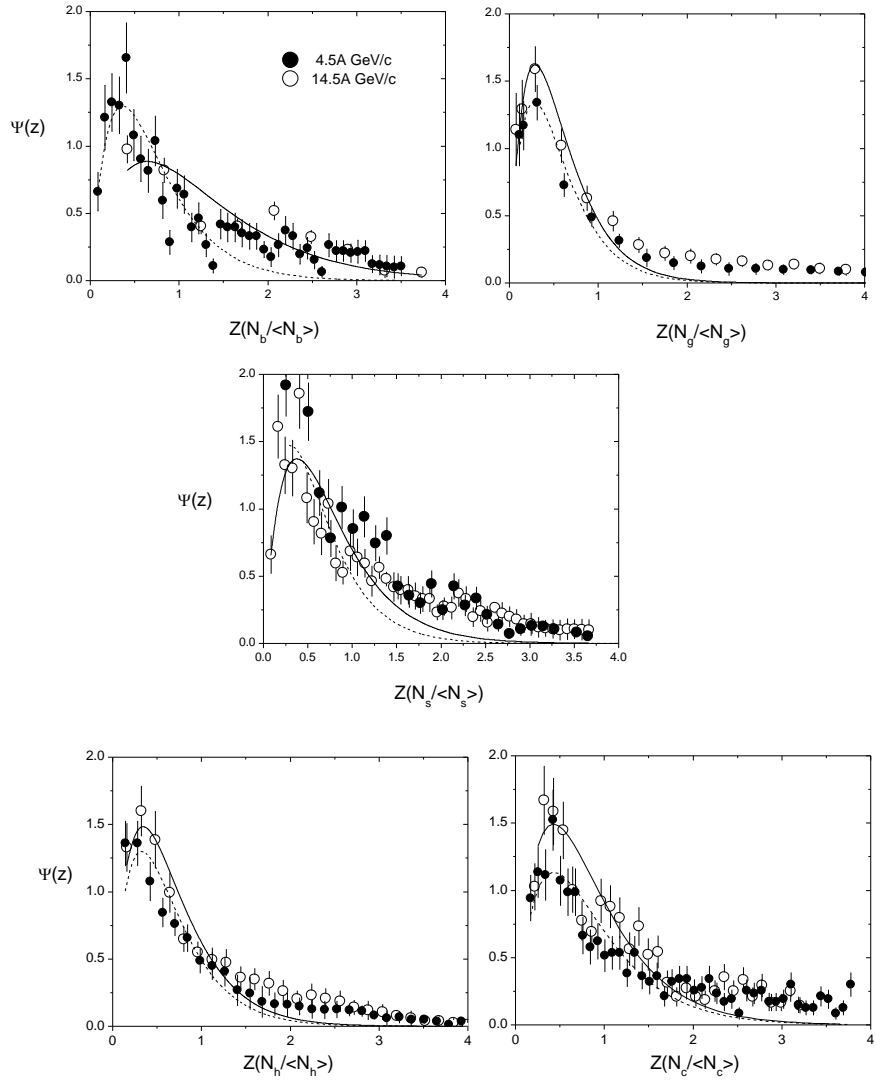

**Fig. 2** Variations of ψ(z) with z for various types of secondary particles produced in $^{28}$Si-nucleus interactions.

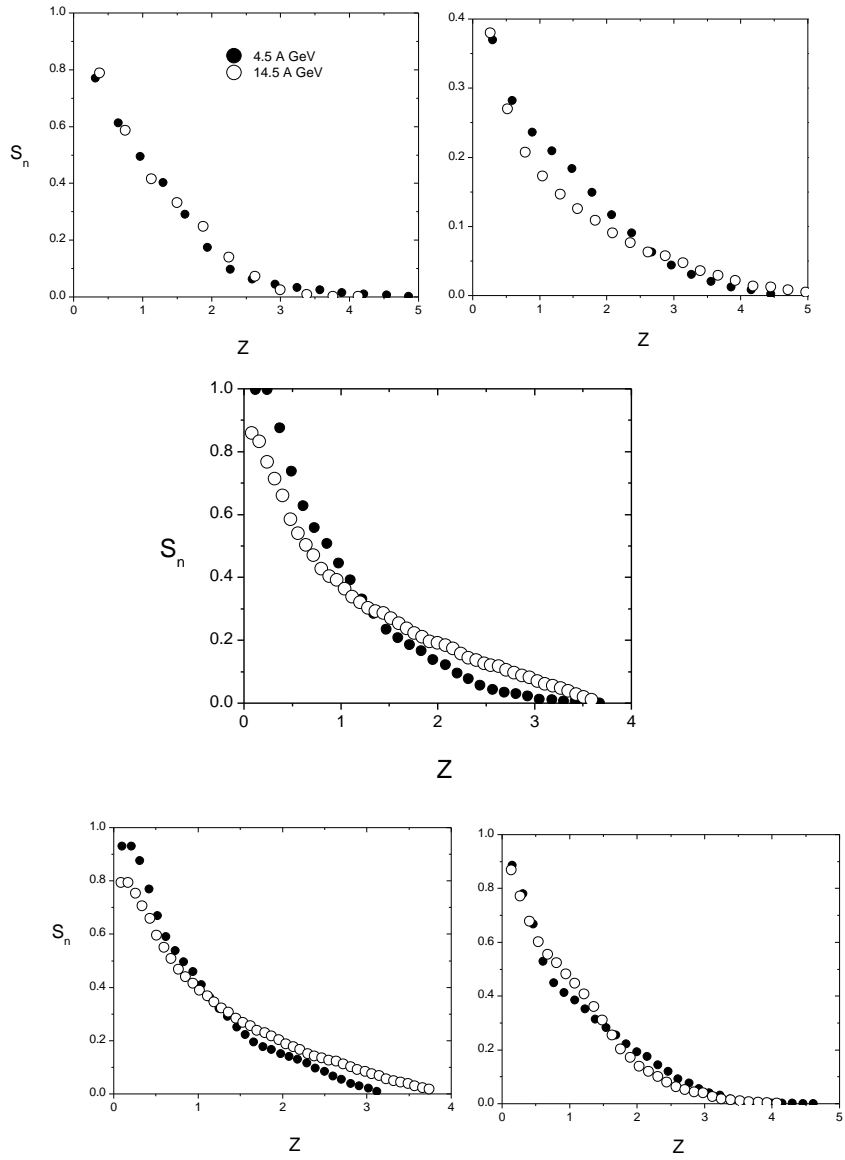

**Fig. 3** Variations of $S_n$ with z for various types of secondary particles produced in $^{28}$Si-nucleus interactions.

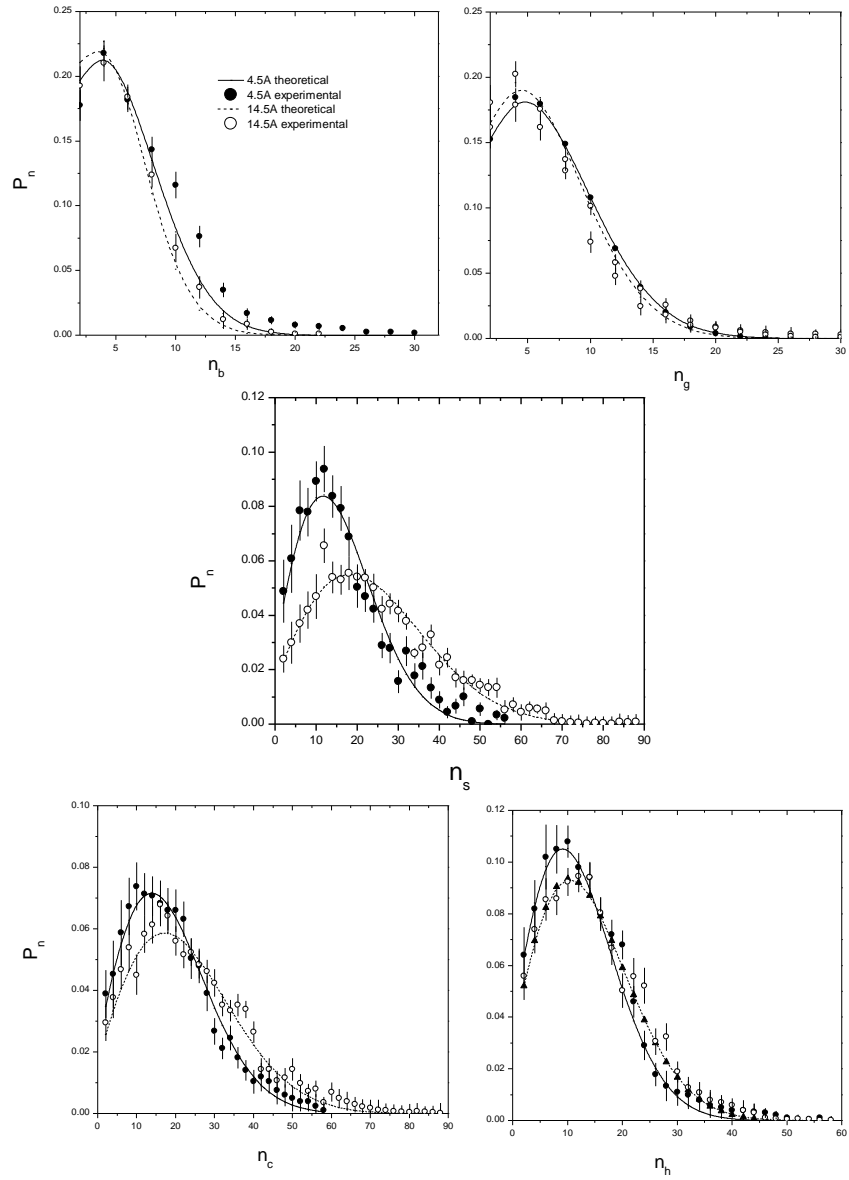

**Fig. 4** Variations of $P_n$ with n for various types of secondary particles produced in $^{28}$Si-nucleus interactions.

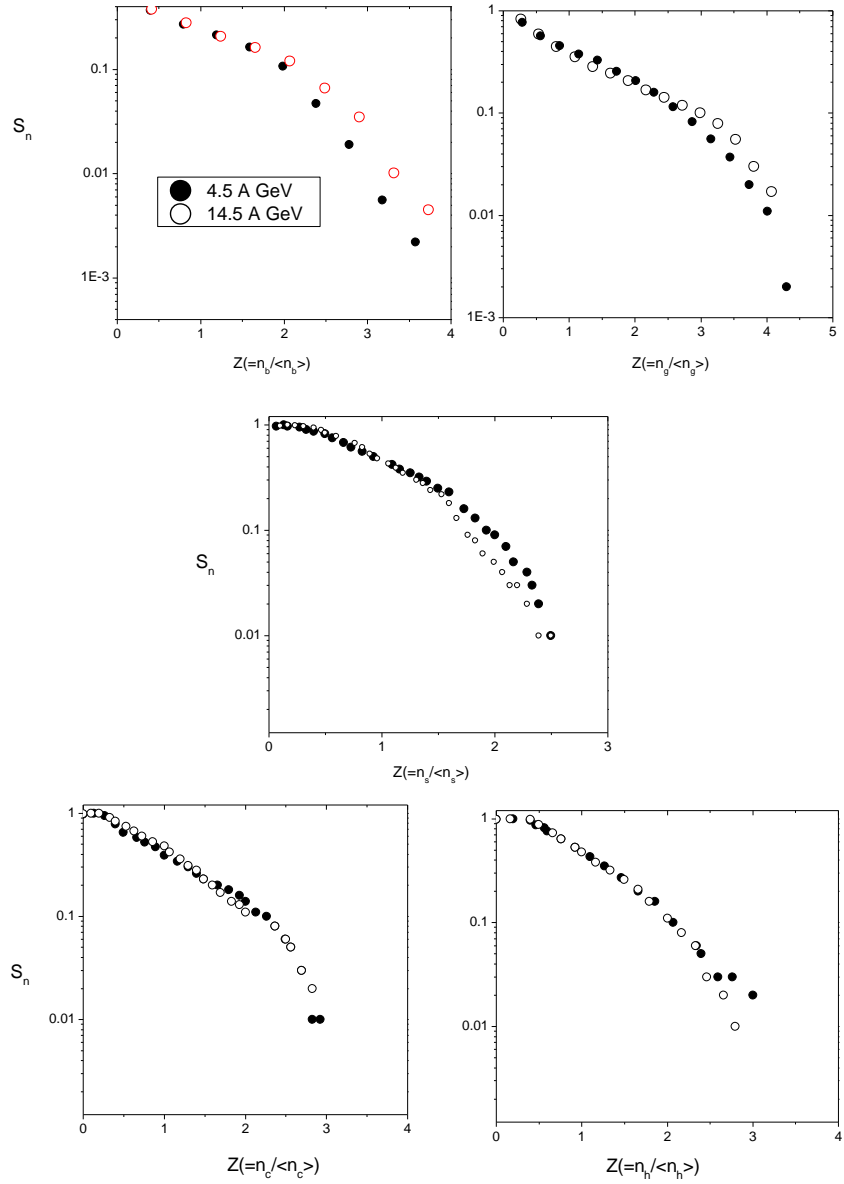

**Fig. 5** Variations of $S_n(z')$ with $z'$ for various types of secondary particles produced in $^{28}$Si-nucleus interactions.

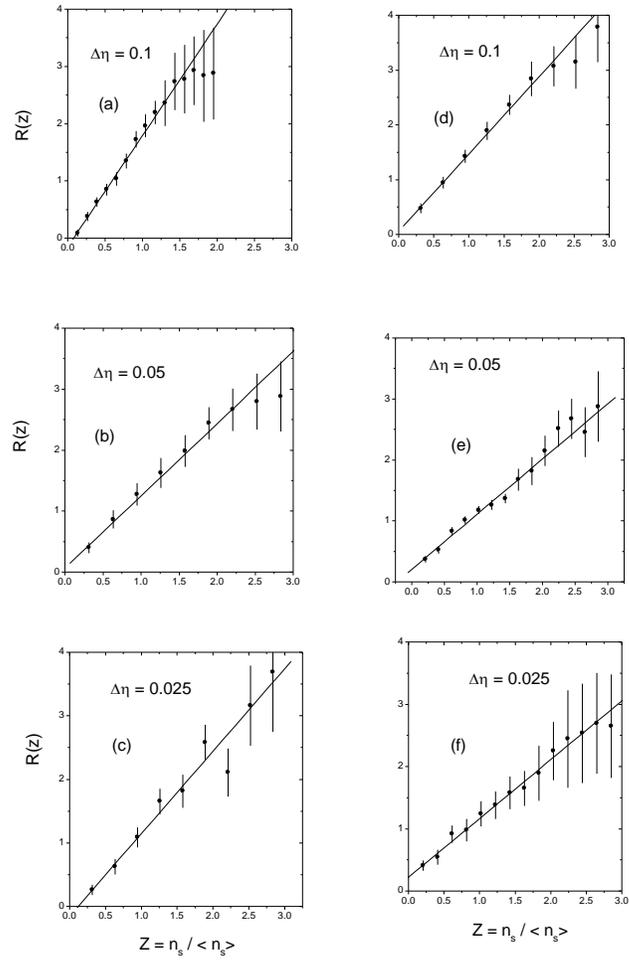

**Fig. 6** Variations of R with Z for 4.5A(a-c) and 14.5A(d-f) GeV $^{28}$Si-nucleus interactions for different rapidity windows.